q-Measurement Scheme set in Quantum Photonic Frameworks:
Linking Abstract Space and Lab Spaces

O. Tapia
Chemistry Ångström, Uppsala University, Sweden

**Summary**
Beginning in abstract space and dislodging the representational form paves a way to formulate a version of a quantum physical measurement scheme. With materiality playing sustainment roles vis-à-vis q-states, these latter control system's response towards external probes. Sustainment concerns all *possible* q-states. The particle character usually assigned to eigen-states and eigenvalues dies out, while energy gaps relate to *possible* energy quanta exchanges; the occupancy concept no longer required. Abstract and laboratory spaces are non-commensurate, linking them lead to a possible physical q-measurement theory. Connectors (links) are referred to as *gangplanks and/or gangways* they perform in frontiers arbitrating quantum to classical regions. Inertial frames facilitate linking q-state domains via projection to wave function space. A physical quantum measurement theory obtains grounded on these elements including quantum information concepts.

**Introduction**
This work examines a possible quantum measurement scheme including actual energy/momentum exchanges. The result goes beyond standard approaches, such as the one exemplified in Chapt.VI of von Neumann treatise. [1] See also ref. [2]

The setting for a quantum measurement theory has changed with inclusion of abstract quantum states that exhibit no representational characters; yet possibly bridged (linked) to both lab q-states (quantum states) and q-information viewpoints. The elementary units of quantum information technologies, quantum bit (q-bit), show basic 2-spinor-like form, with q-states being key elements to uncover their presence. [2-8] At lab level, matter enters via elementary materials *sustaining* q-states; the basis takes on a vector form of infinite dimension: these bases help uncover spectral responses; the scalar product form (include basis and amplitudes) stands for q-states. [9] The standard linear map is not necessarily a faithful presentation of q-states.

For matter-radiation case, photonic frameworks display two regimes of interest: (1) the abstract one, including photon-basis states and matter-sustained ones; and (2) the semi-classic regime, [10] embracing wave functions models and interaction operators to be lodged at a space in-between lab-to-abstract ones.

At lab level, amplitudes relocations among q-partites system require energy quanta exchanges. Lab descriptions go together with angular and/or linear momentum quanta "traded" among them; in this perspective, any q-change affecting interacting systems can be handled via conservation principles. Quantum resonances, on the other hand, provide appropriate means to relate separate partite q-states linked to new (bonded) partites. Finding spectral disparities would signal q-states changes rather than structural modification only; e.g. the activation of spin triplet states starting from related excited singlet spin states. [3,7,8]

Topics required for this bridge construction:
1. Formal QM, the commutator operator e.g., [|i><k|, $H_{op}$] extracts information from a q-state |Φ> that —for diagonal Hamiltonian: $H_{op}$ |i> = $\varepsilon_i$ |i> — leads to:

$$[|i><k| \ H_{op}]|\Phi> = (\varepsilon_i-\varepsilon_k) \ |i><k|\Phi> \rightarrow ih/2\pi \ \partial<k|\Phi>/\partial t \qquad (1)$$

2. The eigen-energies gap ($\varepsilon_i-\varepsilon_k$) offer a possible communication link to an external source, the response strength being gauged by the amplitude <k|Φ>. This latter will open /close address-k to interaction; thence response intensity can be ordered. Moreover, if the gap transforms (or maps) into an energy quantum, a signal could possibly be exchanged. Yet, producing a signal is more sophisticated than this (see below).
3. How does q-scattering formally work in abstract space is examined to sense the differences with standard theory. In this space, the idea of particle trajectories is discarded. It is interaction of q-states with scattering devices the source of new q-states (formalized below).
4. A discussion about context dependence of quantum responses is done with simple models. Sections on coupling probes as recognition mode via interactions allows us introduce radiation-sustained q-states followed by a central section on gangplanks and gangways.
5. Elementary matter-sustained and radiation-sustained q-states lead to entanglement; a simple formalism is presented and discussed below; coherence idea is further examined. A discussion closes the paper.

**Bridging abstract and laboratory levels**
Because the scheme to be developed is non-representational, quantization of classical physics framework is not the path to pursue. Instead, proper interfaces serve to signal in-between abstract to lab spaces. This exclude probing (sensing) relevant q-states in abstract space since these are not material objects; and, in-between regions become necessary; yet, the most one can get is to deliver/recuperate pertinent information. More exactly, in-between states suggest paths to comprehend q-states' roles; e.g.: connections (bridges) among non-commensurate spaces.

To prompt setting domain connections, introduce the concept of inertial frames (I-frames) related to special relativity theory (SRT). I-frames support configuration spaces; for a particular case, the *number* of classical degrees of freedom allows for a dimension assignment, e.g., consider the ket set: {|$x_1$,…,$x_n$>} where coordinates are real numbers (no more) in 3n dimension. Thus two I-frames help outline relative origins (*distance*) and relative orientation (*rotations*) as required by SRT for real space, the requirements on internal elementary materiality express itself only as *presence*; so, coordinates do not ineludibly represent particle positions, they are number sets instead. Time and space introduced on mathematic grounds (Cf. eq.(1) will be assembled at laboratory premises later on. Scalar mappings <$x_1$,…,$x_n$|Ψ> *connect* abstract |Ψ>-states to wave functions, Ψ($x_1$,…,$x_n$); see chapters 1 and 2 in ref.[1] for thorough mathematical analyses. The wave function, in turn, can be used to *link* to complex valued function spaces over real multivalued domain; n-tuples may open *possibilities* for inclusion of semi-classic schemes (e.g. many body systems). [10] Importantly, the label Ψ (the name) is what matters by pointing to particular q-states, of course wavefunctions may open routes to semi-classic frameworks that now, would rather include lab-time; these latter relations require bridges because abstract and lab spaces are non commensurate. The *presence* view (not a location one) of elementary constituents warrants simultaneous existence to all q-states relevant to a physical system (no particles occupation as presence is what matters). A q-state projected as mentioned above is an element never to be taken as an object. Interestingly, I-frames in large numbers would cover real (including) laboratory space; symmetries associated to, enter as gauge symmetries and gauge transformations.

From formal QM, see chapts.1 and 2 in [1], commutator operator e.g., [|i><k|, $H_{op}$] extracts information from a q-state $|\Phi>$, eq.(1) above. The term after arrow sign in (1) implies dynamical aspects not included in its lhs (left hand side); formally, it holds for abstract space (Hilbert space, see end chap.2 in [1]), the time derivative $\partial<k|\Phi>/\partial t$ showing inverse of time dimension. Planck's constant does the job homogenizing dimensionally both sides thereby including a central role for action idea (e,g. energy x time; momentum x distance. For time-independent $H_{op}$, the formal unitary evolution in Hilbert space does not change the amplitudes (in modulus), thus, if lab communications are desired the system must be opened to energy/momentum exchanges; and this would be achieved (or mimicked) in the theoretical scheme; for example, by introducing time dependent operators that prompts for q-states changes. This latter stance calls for a bridge to semi-classic schemes as open partites regions may enter the scene. Thence, time t would enter state ket: $|\Phi, t>$; t is parametric not a dynamic variable, yet, $\Delta t$ (time lapse, interval) may be linked to particular lifetimes or to internal dynamical aspects and not to scalar variable in a time axis; Heisenberg famous uncertainty elements illustrate this point. Time extension may share physical character: e.g. the time lag that takes probing to die out as response, the extension measured in some units.

Yet, energy eigenvalues in (1) are not observable (measurable) in themselves, only energy gaps and amplitudes are. Once this energy quantum is weighted "as it were" by the "emitter" via $<k|\Phi>$-amplitude, the transfer (if any) would imply this overhead information: namely, q-state's information.

*How does q-scattering formally work in abstract space?* The question reminds us that trajectory concept is left behind and that occurrence of real interactions would drive the system out the abstract space (via gangplanks). Acting in a possibility framework, the view of scattering takes on an unusual facet: there are no particles interacting with a device, but *q-states do*. Thence, one should be able to sense possibilities shaped by interacting centers; *interaction operators* ($V_\wedge$) act on basis q-states. Operator $V_\wedge$ is presented in practice by way of semi-classic elements where lab-location tags scattering centers. The effects calculated from resulting $V_\wedge|\Psi>$-states come out all at once yet, they are not "caused" by sustaining element in any particular manner. The connection can be seen as a relational one. However, advanced and retarded modes are effective only at the in-between and lab regions. *Classical causality is superseded because QM is describing no real material elements (objects); it only concerns q-states*.

In summary, *q-scattering produces new q-states*: Amplitudes e.g. $\{<k|V_\wedge|\Psi>\}_k$, would express all possible scattering directions counted via base states. The which-way pictures ("particle") become irrelevant (a classical physics remnant only); the key being: uncovering (production) of new q-state(s). At lab-level there are numbers of specific devices that symbolically may take the place of $V_\wedge$-operators: few examples are: i) beam splitters (BSs), simple and/or multiport BSs; ii) operator related to non-linear crystal (NLC) leading to entangled states, etc. Note, configuration space is a mathematical construct not concerned with particle representations or dynamics.

The amplitude $V_\wedge|\Psi>$ brings in an encircling information; that, *when linked* to a lab-view, can be *styled* by graphs "blue-prints" (schemes) drawing relevant ray-directions; these rays might *seemingly* be seen as emerging from (or sinking at) scattering centers (or probing devices), they decorate a design. When several other devices are present these rays can cross at specific lab space points; all of them taken as *guiding helps* and, of course, by no means standing as particle trajectory representations. The corresponding drawing emphasizes a *technology dimension*.

Even if, on the one hand, q-states do not represent elementary materiality, on the other hand, particular lab-arrangement would help breed knowledge (information) about the system's *response patterns*, which turns out to be a key communication element. Actual drawings and response patterns may turn explicit our theoretical expectations.

In summary, there is no quantum description of putative physical reality; what there is, relate to probing associated to a mathematical physics project.

Illustration: take operators $V_{\wedge 1}$ and $V_{\wedge 2}$ each linked to different lab scattering centers, and form $(V_{\wedge 1} \oplus V_{\wedge 2})|\Psi>$ where operators generate new q-states namely: $|\Psi_{V\wedge 1}>$ for $V_{\wedge 1}|\Psi>$ and $|\Psi_{V\wedge 2}>$ for $V_{\wedge 2}|\Psi>$, respectively. In principle, one can use any lab guiding directions to *gather relevant detailed information*; when located at lab space, appropriate probing centers may operate. [7,15] Yet these directions would serve locate amplitude' sources (e.g., galaxies). When ray intersections are encountered, new equipment, intervening in a supervenient (extrinsic) manner, can locally help probing scalar-products-states; eventual results can be detected at in-between domains with appropriate devices (sensors at lab space).

This situation brings us into a known case of Contextuality mode. A context is one of the elements that would help organize knowledge about systems: it has nothing to do with the putative location of elementary matter sustaining the q-states.

*Introducing probes to help recognition interactions (measurement)*
To construct a more suitable formalism, take radiation- and matter- sustained q-states separately prepared.[7-9] The functionality expressed via q-states involve a direct sum: e.g., |*radiation-sustained*> ⊕ |*elementary-matter-sustained*>.

Logical sum (⊕) states [6] signal non-interaction and helps gathering information conveying data on global elementary material elements and/or radiation types. Consider *radiation-sustained q-states* first: The 2-form standing for 1-photon base state, qbit, reads:

$$|1\text{photon basis}; \omega> \rightarrow (|0_\omega> \quad |1_\omega>) \qquad (2)$$

Fock space labels $0_\omega$ and $1_\omega$ refer to photon-number model where the basis links up a colored vacuum to one-photon elements. A q-state *sustained by radiation energy* maps to Hilbert space component: $(C_{|0_\omega>} \quad C_{|1_\omega>})^t$; with amplitudes $C_{|0_\omega>}$ and $C_{|1_\omega>}$ that are *labeled* complex numbers. Information aspect supersedes classical physics representational one (wave packet). A one-photon q-state ($|$ q-bit, $\omega >$) links to the form: $<$1photon basis$|$ q-bit, $\omega > \rightarrow (|0_\omega> \quad |1_\omega>) \bullet (C_{|0\omega>} \quad C_{|1\omega>})^t$ (3)

In terse notation, the quantum state can simply be mapped onto form:

$$|q\text{-state}, \omega > \rightarrow \quad (C_{|0\omega>} \quad C_{|1\omega>})^t \quad (3')$$

The map (3') is valid only if basis (2) is kept fix and actually stands for eq.(3). This form links to a Hilbert space element; no one-photon "particle" representation implied. For the q-state, normalization gets: $|C_{|0\omega>}|^2 + |C_{|1\omega>}|^2 = 1$; consequently, these amplitudes do not change independently: for *isolated* systems only, $(|C_{|1\omega>}|^2 - 1)$ equaling $|C_{|0\omega>}|^2$. Response in intensity from $|1\omega>$-level is gauged by $|C_{|1\omega>}|^2$ yet it does not indicate a partially populated energy level since a q-system has no properties such as occupancy (except for semi-classic models).

Possible processes involving quanta are expressed via amplitude variations only. The linear form, usually associated to (3), is a map from scalar product to a scalar function;[11,12] yet this is not "representation" of $|q\text{-bit},\omega >$ in spite of its broad use as such;[11-17] moreover, nothing is fluctuating "at frequency $\omega$" because the present approach goes beyond classical Maxwell electromagnetic (EM) theory; currently $\omega$ would reckon as an *energy amount* via Planck's constant yet not as a "particle".

*Gangplanks & gangways*

Consider symbol $(0_{|0\omega>} \quad 0_{|1\omega>})^t$, it can play linking roles as idle channel definition, e.g., shifting from q-state $(0_{|0\omega>} \quad 1_{|1\omega>})^t$ via $(0_{|0\omega>} \quad 0_{|1\omega>})^t$ would mean that an energy quantum can be *displaced* from a radiation field, leaving behind $(1_{|0\omega>} \quad 0_{|1\omega>})^t$ as hole state; or a matter linked one (see below). While in the opposite direction: displacement using the gangplank $(0_{|0\omega>} \quad 0_{|1\omega>})^t$ linking to $(0_{|0\omega>} \quad 1_{|1\omega>})^t$ conveys "transfer" possibility towards photon fields, the labels ensure information able-ness (not representation though implicit transit). The link idea also permits joining two different I-frame sustained states, e.g.:

$(0_{|0\omega>} \quad 1_{|1\omega>})^t{}_A \times \{(0_{|0\omega>} \quad 0_{|1\omega>})^t{}_{gangplank}\}_{AB} \rightarrow$ May possibly put 1quantum in $_B$. (4)

Read in opposite direction it signals a *possibility* for quantum in B transferred to A via gangplank. Read from left to right, possibility to leave a hole in A: $(1_{|0\omega>} \quad 0_{|1\omega>})^t{}_A$. On the contrary,

$(1_{|0\omega>} \quad 0_{|1\omega>})^t{}_A \times \{(0_{|0\omega>} \quad 0_{|1\omega>})^t{}_{gangplank}\}_{AB} \rightarrow$ May possibly put 1hole in $_B$. (5)

Not that the examples emphasize *possibility* for transport from one space to another not a mechanical event; the gangplank bridges the two ends: e.g., from (4) to (5). In this manner, the quantum physical characteristic is not obliterated.

The in-between space hosting a gangplank or gangway, would mediate such events or else the situation remains possible only (no actualization) and the quantum physical character of the situation ensured. Classic and quantum are irreducible.

However, if such event transmute into actual energy, a gangplank will necessarily mediate it. Thus, on the one hand, this option would be a process taking a q-state out from abstract space; and consequently unitary evolution stops! On the other hand, "post mortem" it results in an *information erasure* and consequently thermodynamic entropy of the world changes. This is a reason that justifies an arrow connection only in eq.(1) that otherwise would break down.

*Nevertheless, what really matters is that q-events can be registered and/or saved* [9] *and this saving with appropriate labels can be used for further studies* e.g., imaging analyses, and/or engineering the impulses to be used elsewhere; in one word, q-technology becomes present.

Circumstances of this nature ground a type of measurements in Quantum physics leading to piecewise knowledge and understanding. This later shares no similarity with classical physics one. The concept of gangplank/gangway play central roles underlining presence of *non-commensurate* spaces so that simple transit from one to the other is not a resource; below it will be looked at in more detail.

*Elementary matter-sustained q-states and* r*adiation-sustained ones*
The matter-sustained q-state concept supplemented with radiation sustained ones permit managing radiation-matter interactions beyond the classical physics and semi-classic levels.

As illustration, ponder two independent *partite states* [6,7] e.g. 2-form and a standard Hilbert space vector:

$$(0_{|0_\omega>} \quad 1_{|1_\omega>}) \oplus (1_{|i=0>} \quad 0_{|i=1>} \ldots 0_{|i=2>} \quad 0_{|i=3>} \ldots 0_{|i=n>} \ldots) \quad (6)$$

Basis set ordered with labels patterns including interactions are not conveyed thru $\oplus$-operations. Instead, it is tempered via direct product $\otimes$-forms so that interactions enter in the formalism. Consider the direct product: symbols $\oplus$ changed by $\otimes$ in (6):

$$(0_{|0_\omega>} \quad 1_{|1_\omega>}) \otimes (1_{|i=0>} \quad 0_{|i=1>} \ldots 0_{|i=2>} \quad 0_{|i=3>} \ldots 0_{|i=n>} \ldots) \quad (7)$$

Yet, the basis vector should include possibilities that are not necessarily present here. This is so because *entanglement* enters stage; and the situation changes noticeably. An extended basis vector illustrating such new possibilities would look like (8):

$$(|i=0>\otimes|0_\omega> \quad |i=0>\otimes|1_\omega> \quad |i=0;0_\omega> \quad |i=0;1_\omega> \quad |i=1>\otimes|0_\omega> \quad |i=1>\otimes|1_\omega>$$
$$|i=1;0_\omega> \quad |i=1;1_\omega> \ldots |i=n>\otimes|0_\omega> \quad |i=n>\otimes|1_\omega> \quad |i=n;0_\omega> \quad |i=n;1_\omega> \ldots) \quad (8)$$

In (8) each spinor element make up as slots and q-states would looks like:

$$( C_{i=0\otimes 0\omega} \quad C_{i=0\otimes 1\omega} \quad C_{i=0;0\omega} \quad C_{i=0;1\omega} \quad C_{i=1\otimes 0\omega} \quad C_{i=1\otimes 1\omega} \quad C_{i=1;0\omega} \quad C_{i=1;1\omega} \ldots )^t \quad (9)$$

This stands for a fully entangled radiation-matter q-state, namely a coherent q-state. Thus, besides standard direct product elements there are slots such as $|i=n;0_\omega>$ and $|i=n;1_\omega>$, n=0,1…; the extra niches stand for elementary full entanglement possibility for which spinor state elements and photon state labels are disguised.

This type of basis vector permits handling possibilities for matter-radiation interaction q-states denoted with labeled amplitudes as in vector form (9); note, once a choice of basis states is made, they must remain fixed for given measuring processes only amplitudes detecting dynamic effects. An example, take q-state reshuffled as:

$$( 0_{i=0\otimes(0\omega \; 1\omega)} \quad 0_{i=0;(0\omega \; 1\omega)} \quad 0_{i=1\otimes(0\omega \; 1\omega)} \quad 1_{i=1;(0\omega \; 1\omega)} \ldots 0_{|i=n>\otimes(0\omega \; 1\omega)} \; 0_{|i=n>;(0\omega \; 1\omega)} \ldots ) \quad (10)$$

The label in $1_{i=1;\,(0\omega \; 1\omega)}$ suggest a two-photon case situation, namely, the one required to shift the label from i=0-to-i=1 and the second 1-photon state "melted" as it were in the sustaining materiality and, consequently, not accessible as such. In simple words, non-linear situations are then in principle available possibilities; the advantage is that photon slots with 2,3…n- quanta can be accommodated as possibilities. Any entanglement change would be achieved via quantum physical processes only corresponding to resonances for partites entering as incoherent states elements.

Via algebraic operations neither radiation-sustained nor matter-sustained elements can be separated from a coherent form like (9) or specially (10), therefore no simple mechanical (algebraic) models would allow to do this; one is confronted with "uncontrollable" disturbances, were we measuring it. Yet, in the present approach each particular slot base-vector shows information on *possible* interaction *response* patterns as function of pertinent amplitudes (Cf.eq.1), the whole characterizing a complex response situation yet never a collapse of the wave function. Any *naught amplitude closes access to that specific interaction* yet its position in the state vector remains, thereby retaining possibilities that otherwise would be erased, e.g. in echoing. This instance differs from linear standard models and marks irreducibility and a kind of novelty.

Note that for each energy level a fourfold degeneracy become apparent in (9); care is required to identify emission/absorption possibilities. As basis set (8) indicates possibilities must be reckoned (in advance), they are not "produced" while interaction is going on (so to speak); this disagrees with classical physics views, completely.

How does a gangplank work in emission/absorption processes?

$$(1_{|0\omega>} \quad 0_{|1\omega>}) \otimes (0_{i=0 \otimes 0\omega} \; 0_{i=0 \otimes 1\omega} 0_{i=0;0\omega} 0_{i=0;1\omega} \ldots 1_{i=1 \otimes 0\omega} 0_{i=1;0\omega} 0_{i=1 \otimes 1\omega} \ldots)^t \quad (11)$$

A connection opening a channel to (11) corresponds to a gangplank: $(0_{|0\omega>} \quad 0_{|1\omega})$ unlocking possibilities for inclusion of ingoing or outgoing energy quantum episodes. Observe that the transit of energy quanta can occur in lab or in real space not in abstract space. Amplitudes in Hilbert space can be tag with information only. The channel associates now to a non-standard operation (×) pointing action to *gangplank connectors*:

$$(0_{|0\omega>} \quad 0_{|1\omega}) \times (0_{i=0 \otimes 0\omega} \ldots 0_{i=0 \otimes 1\omega} \; 0_{i=0;0\omega} \; 0_{i=0;1\omega} \ldots 1_{i=1 \otimes 0\omega} \; 0_{i=1;0\omega} \ldots)^t \quad (12)$$

$$(0_{|0\omega>} \quad 1_{|1\omega}) \times (0_{i=0 \otimes 0\omega} \ldots 0_{i=0 \otimes 1\omega} \; 0_{i=0;0\omega} \; 0_{i=0;1\omega} \ldots 0_{i=1 \otimes 0\omega} \; 0_{i=1;0\omega} \ldots)^t \quad (12')$$

In (12), an energy quantum activates the excited state slot at gangplank region; and now it may possibly leave the radiation field. While the possibility signaled by (12') stands for the opposite effect. It is matter-sustained that would act as ingoing channel state. Both (12) and (12') are present so that a prediction in a classical sense is forbidden thereby uncovering a quantum character at this level too. The technological freedom experimentalists may have to arrange measurements (probing) devices is implied by ×-operation — i.e., context imposing. The "result" rooted at (12)-(12') shares quantum physical character even if the outcome *would* (materially) affect either the radiation or the matter fields. Thus, theoretically one may start up transfer towards a radiation field but *only as a possibility*; both steps taken together display a sort of "Schrödinger-cat" q-state [2] that effectively locates here at an in-between space. The connector would correspondingly share quantum physical character and consequently, predictive capabilities (in the classical mechanical sense) are not available, ever. The outcomes in time and space share a random quality not statistical.

Also, were an irreversible process occur, the point-result constitutes one way to collect information about a q-state entering (1). Thus, gangplanks share an important quantum physical character acting as bridges between quantum physical states. Symbols $(0_{|0\omega>} \quad 0_{|1\omega})$ or $(0_{i=0\otimes 0\omega} \ldots 0_{i=0\otimes 1\omega} \quad 0_{i=0;0\omega} \quad 0_{i=0;1\omega} \ldots 0_{i=1\otimes 0\omega} \quad 0_{i=1;0\omega} \ldots)^t$ justify in a way the talk on idle q-states (gang-planks).

*Interference with Independent Photon Beams*

The context, tips off a pair of 2-dimensional generic bases, one linked to laser-1(A) the other to laser-2(B) in parallel: $(|\Xi_{SAx}> \quad |\Xi_{SAy}>)$ and $(|\Xi_{SBx}> \quad |\Xi_{SBy}>)$, respectively. Note, in this case basis vectors are always parallel translated wherever a q-state is to be specified (yet they do not transport as if they were particles).

A diagram *styled* drawing ray-directions from laser sources looks as follows: Laser-1(A), ray-1 along x-direction encounter mirror $M_1$, the interaction $V_{M1}|\Xi_{SA}>$ generating two q-states; one, along y-axis in $M_2$-direction (i.e., orthogonal to x), the second, continues along x, pointing in direction to mirror $M_4$. First, directly link $M_1$-to-$M_4$ or close a loop via $M_1$-$M_2$-$M_3$-$M_4$ thus two incoming coherent q-states at M4. Mirrors $M_1$, $M_2$ and $M_4$ are half silvered, $M_3$ is standard.

Can one obtain coherence involving independent laser sources? The answer in this framework is negative as q-states have different roots.

Let us examine this issue. The amplitudes associated to $V_{M1}|\Xi_{SA}>$ and $V_{M2}|\Xi_{SB}>$ bring in encircling information when the four mirrors set up are in place. Let search further: Laser-1 runs along loops $M_1$-$M_2$-$M_3$-$M_4$ (rectangular) and $M_1$-$M_4$ (direct) the *basis set* is conserved, namely $(|\Xi_{SAx}> \quad |\Xi_{SAy}>)$; while for Laser-2(B) the base is obviously different: $(|\Xi_{SBx}> \quad |\Xi_{SBy}>)$; and no link is present to mixing different basis sets states. A simple calculation shows how does it work the abstract-to-lab framework. The q-state $(C_{|\Xi S1x>}=1 \quad C_{|\Xi S1y>}=0)^t$ to be scattered at $M_1$ (source states $S_1$) while if source $S_2$ is selected, generic input q-states reads $(C_{|\Xi S2x>}=1 \quad C_{|\Xi S2y>}=0)^t$ that are styled in parallel lab directions but referring to different elementary material sustaining systems, e.g., different frequency and/or phases. Use labels fixing the source origin for scattered q-states:

1-A) Along direction 1-2 → $(C_{|\Xi S1x>}= 0 \quad C_{|\Xi S1y>}= -i/\sqrt{2})^t_{1-2}$ $C_{1-2}(1)$ (y-direction down)
2-A) Along direction 1-4 → $(C_{|\Xi S1x>}=1/\sqrt{2} \quad C_{|\Xi S1y>}=0)^t_{1-4}$ $C_{1-4}(1)$ ( x-direction)
3-A) Along direction 2-3 → $(C_{|\Xi S1x>}=1/\sqrt{2} \quad C_{|\Xi S1y>}=0)^t_{2-3}$ $C_{2-3}(1)$ (x-direction)
4-A) Along direction 3-4 → $(C_{|\Xi S1x>}=0 \quad C_{|\Xi S1y>}= i/\sqrt{2})^t_{3-4}$ $C_{3-4}(1)$ (y-direction up)

For source 2 output $(C_{|\Xi S2x>}=1 \quad C_{|\Xi S2y>}=0)^t$ select after $M_2$:

1-B) Along direction 2-1 →$(C_{|\Xi S2x>}=0 \quad C_{|\Xi S2y>}=+i/\sqrt{2})^t_{1-2}$ $C_{2-1}(2)$ (y-direction up)
2-B) Along direction 2-3 →$(C_{|\Xi S2x>}=1/\sqrt{2} \quad C_{|\Xi S2y>}=0)^t_{2-3}$ $C_{2-3}(2)$ ( x-direction)
3-B) Along direction 3-4 →$(C_{|\Xi S2x>}=0 \quad C_{|\Xi S2y>}=1/\sqrt{2})^t_{3-4}$ $C_{3-4}(2)$ (y-direction up)
4-B) Along direction 1-4 →$(C_{|\Xi S2x>}=1/\sqrt{2} \quad C_{|\Xi S2y>}=0)^t_{1-2}$ $C_{1-4}(2)$ ( x-direction)

And, e.g. $C_{1-4}(2) = 0$ would mean, by construction, no interferences expected to be rooted in laser-2. Of course, this latter condition belongs to experimenter's decision thereby reflecting the probing character of the device. With assignments:

$\phi_{1-4} \rightarrow (C_{|\Xi S1x>}=1/\sqrt{2} \quad C_{|\Xi S1y>}=0)^t_{1-4}$ and $\phi_{3-4} \rightarrow (C_{|\Xi S1x>}=0 \quad C_{|\Xi S1y>}=1/\sqrt{2})^t_{3-4}$.

And for a generic situation, with $C_{1-4}$ & $C_{3-4}$ equal unity, the q-state to be probed takes on the form $\phi_{1-4} + \phi_{3-4}$ that in intensity regime: $|\phi_{1-4} + \phi_{3-4}|^2$ show possibilities for interference that eventually could be transformed into q-events.

Thus, if interference there were, it would involve the same elementary sustaining materiality thus same source. Independent photon sources do not interfere with this arrangement type.

**Discussion**
This work requires QM formalism to remain intact in abstract Hilbert space, yet when grounds shift from probability-to-possibility mode this implies changing the interpretive context and, as a result, superseding standard measurement approach. [1,2]

After imposing distinctions between abstract and laboratory space the modified wit wipe out Quantum Mechanics' weirdness: no particle and no paths. Second, the approach prompts for links (connections) between otherwise non-commensurable spaces. [1,2,9] It follows that abstract quantum theory alone shouldn't be used to examine laboratory-based events; at boundaries unitary time evolution breaks inevitably; for the simple reason that lab-system corresponds then to an open one, thus breaking conditions to properly implement Hilbert space mathematical varieties.

From q-scattering perspective, *new interaction-generated q-states open fields over new possibilities*, renewed information circulates E.g., double-slit experiments; see pages 60 to 78 ref. [9] and once you realize that full sets of q-states become possible then accessibility must be granted *in the theoretical framework*. Note that scattering effects are not due to simple passage of particles following putative real space trajectories. Uncovering of interference patterns firstly becomes a mathematical result generated by q-states superposition, e.g. $|\phi_{1-4} + \phi_{3-4}|^2$ thereafter focus move to laboratory space; and in-between spaces are now central. [8,9,18]

Radiation q-state carries information on both, energy and angular momenta (spin), and it opens paths to implement entanglement patterns within the photonic scheme. Basis state kets structure, given by eq. (7), (8), signals coherent domain. [5-8]

The links one could construct between non-commensurate spaces were identified with gangplanks. Moreover, via *Feshbach resonances* [13] non-commensurate spaces can also be linked, and only when these resonances signal particular gangplank cases the link becomes fully quantum physical. All these structures act in possibility mode unless actual energy-momentum transfer takes place; in this latter case, it would prompt for irreversible thermodynamic effects (e.g. Landauer effect). [19]

Initiation of a measurement process necessarily imposes a partitioning step into non-commensurable partites. Thus, new connecting space (gangplank) is required so that signal registering finds a proper sustainment after prompting q-events occurrence; yet, timing and space locations are not predictable. Even if q-events at lab were recorded one-by-one (as they usually are), each "spot" would always include amplitude information (Cf. eq. (1). Questions such as: which way a "photon" or an "electron" took, for instance at double-slit or BS, *are empty statements*. A reason is that such accounts do not belong to a quantum description; q-states do not relate directly or indirectly to real space trajectories; such type of questions belongs to classical physics world problems; [13-15] and these cases are not accountable by quantum schemes, [15,16] also read note 148 ref.[1]; we partially quote it from page 283: [1] *"It has often been said that the quantum mechanics involves …dual nature, since the discrete particles (electrons, protons) are also described by wave functions, and exhibit typical wave properties…In contrast with this,…quantum* *mechanics derives both "natures" from a single unified theory of elementary phenomena*"; we add: only q-states actually count but not waves or particles in their classical sense; these latter ideas distortion quantum descriptions.

The bridging of non-commensurable spaces can also be achieved resourcing to procedures involving *entanglements to partially re-connect them*. The partites' elementary materials integrate the new form (entity word is not adequate) yet the initial partites q-nature fades away as they do not reappear as such in the global system: e.g. the hydrogen molecule does not "contain" two hydrogen atoms. Note, bridge-states acquire partial quantum character including *idle photon states* that also become instrumental as *linkers* between abstract-lab domains, e.g., (12) & (12').

"*The primary lesson taught by quantum theory is that the structure of empirically observed macroscopic phenomena cannot be understood within a conceptual framework, in which the course of physical events is determined by local mechanical laws of the kind specified by laws of classical physics*" Cf. H.P. Stapp (ref. [19] to be found there as paper Nb.20). The same conclusion follows from the present work.

Here, handling of measurement shows that there is no requirement for observers, only recorders and interpreters (be it human or/and artificial-intelligence-based machines). Bridges and links are the elements integrating a theory of measurements. Measuring/detecting atomic spectra first identify the source of radiation and thence set up a probe intersecting the ray linking both elements (e.g., spectroscope).

The fact that eigenvalues are not individually measurable does not detract the mathematical structure of QM. They help introduce order relationships independently of any interpretive scheme. See ref. [20] for a contemporary mathematical presentation.

Finally, gangplanks and gangways always display quantum physical behavior in "connection" with quantum systems. This is why the vector form is used to discuss their roles. In addition, there would never be a simple classical link between a q-system and a laboratory classic one. The language, however, remains historically rooted in classical world practices; [4,7-9] the classical view of photon and electron transfer is ignored in abstract QM although semi-classic (objective) QM [21] may partly recover it including representational elements; but this is another story. [22]

**Acknowledgment**

# References

1. J. Von Neumann, *Mathematical foundations of quantum mechanics* (Princeton University Press, 1955)
2. J.D. Trimmer, *Present Situation in Quantum Mechanics: A translation of Schrödinger's "Cat Paradox" Paper*, Proc. Am. Philos. Soc. 124, 323-338 (1980)
3. O. Tapia, Photonic Framework: *Cues To Decode Quantum Mechanics*. (2017) arXiv:1710.01514v1 [quant-ph]
4. O. Tapia, *Quantum Entanglement and Decoherence: Beyond Particle Models. A Farewell to Quantum Mechanic's Weirdness*. (2014) arXiv:1404.0552v1 [quant-ph]
5. O. Tapia, *Quantum-matter photonic framework perspective of chemical processes: Entanglement shifts in HCN/CNH isomerization.* Int.J. Quantum Chem.115, 1490 (2015)
6. O. Tapia, *Quantum photonic base states: concept and molecular modeling. Managing chemical process descriptions beyond semi-classic schemes.* J. Mol. Model 20, 2110 (2014)
7. O. Tapia, *Quantum Physics: From Abstract to Laboratory Space I. Q-States Sustained by Partite Material Systems: Linking $A \oplus B$ and $A \otimes B$ domains via Entanglement*. (2014) arXiv:1706.00288v1 [quant-ph]
8. O. Tapia, *State-quantum-chemistry set in a photonic framework. Adv. Quantum Chem.* 74, 227-251 (2017)
9. O. Tapia, *Quantum states for quantum measurements*, Adv. Quantum Chem. 56,31 (2009)
10. Crespo R, Piqueras M-C, Aullo J.M, Tapia O, *A Theoretical Study of Semiclassic Models: Toward a Quantum Mechanical Representation of Chemical Processes.* Int.J. Quantum Chem. 111, 263-271 (2011)
11. R.J. Glauber, *Coherent and Incoherent States of the Radiation Field.* Phys.Rev. 131, 2766-2788 (1963)
12. 21. D. Markham, V. Vedral. *Classicality of spin coherent states via entanglement and distinguishability*. Phys.Rev. A 67, 042113 (2003)
13. G.A Arteca, Aulló J.M., O. Tapia. *Constructing quantum mechanical models starting from diabatic schemes: I. Feshbach-like quantum states and electronuclear wavefunctions.* J.Math.Chem. 50, 949-970 (2012)
14. A.J.Leggett, *Macroscopic Quantum Systems and the Quantum Theory of Measurement.* Supp.Progr.Theoret.Phys. No.69, 80-100 (1980)
15. Tapia Olivares Orlando (2012) *Quantum Physical Chemistry*, Publicacions Universitat Jaume I, ISBN 978-84-8021-828-3, Castelló, Spain
16. K-H Song, et al.*Quantum Beats and Phase Shifts in Two-Dimensional Electronic Spectra of Zinc Naphthalo-cyanine Monomer and Aggregate*, J.Phys.Chem.Lett. 6 (2015) 4314-4318
17. J.Kincaid, McLelland K., Zwolak M. *Measurement-induced decoherence and information in double-slit interference*. Am.J.Phys. 84, 522 (2016); doi: 10.1119/1.4943585
18. M.B. Plenio, V. Vitelli. *The physics of forgetting: Landauer's erasure principle and information theory*. (2001) http //arXiv: quant-ph/0103108v1
19. A. Elitzur, Dolev S., Kolenda N.(Eds.) *Quo Vadis Quantum Mechanics?* (Springer, 2005)
20. F. Strocchi. Mathematical Structure of Quantum Mechanics (World Scientific 2005) Adv. Series Mathematical Physics - vol.27.
21. F.J. Belifante. *Measurements and time reversal in objective quantum theory*. (Pergamon Press,1975)
22. A. Gover, Yariv A. *Free-electron-Bound-Electron resonant interaction*. Phys.Rev.Lett. 124 (2020) 064801